\documentclass[12pt,preprint]{aastex}
\begin{document}
\title{Reconstructing the Intrinsic Triaxial Shape of the Virgo Cluster}
\author{Bomee Lee and Jounghun Lee}
\affil{Department of Physics and Astronomy, FPRD, Seoul National University, 
Seoul 151-747, Korea} 
\email{bmlee@astro.snu.ac.kr, jounghun@astro.snu.ac.kr}
\begin{abstract}
To use galaxy clusters as a cosmological probe, it is important to account 
for their triaxiality. Assuming that the triaxial shapes of galaxy clusters 
are induced by the tidal interaction with the surrounding matter, Lee and 
Kang recently developed a reconstruction algorithm for the measurement of 
the axial ratio of a triaxial cluster. We examine the validity of this 
reconstruction algorithm by performing an observational test of it with 
the Virgo cluster as a target. We first modify the LK06 algorithm by 
incorporating the two dimensional projection effect. Then, we analyze 
the $1275$ member galaxies from the Virgo Cluster Catalogue and find 
the projected direction of the Virgo cluster major axis by measuring the 
anisotropy in the spatial distribution of the member galaxies in the two 
dimensional projected plane. Applying the modified reconstruction algorithm 
to the analyzed data, we find that the axial ratio of the triaxial 
Virgo cluster is ($1$: $0.54$ : $0.73$). This result is consistent with the 
recent observational report from the Virgo Cluster Survey, proving the 
robustness of the reconstruction algorithm. It is also found that at the 
inner radii the shape tends to be more like prolate. We discuss the 
possible effect of the Virgo cluster triaxiality on the mass estimation.
\end{abstract}

\section{INTRODUCTION}

Galaxy clusters provide one of the most powerful tools to constrain the key 
cosmological parameters. In the era of precision cosmology, it is 
important to determine their mass as accurately as possible before using 
them as a cosmological probe. Any kind of simplified assumption about 
the properties of galaxy clusters could cause substantial systematics 
in the mass estimation. The triaxial shapes of galaxy clusters are one 
of such properties. 

It has been long known both observationally and numerically that the galaxy 
clusters are noticeably triaxial \citep[e.g.,][]{fre-etal88,wes89,plio91,
war-etal92}. While plenty of efforts have been already made to take into 
account the triaxial shapes of galaxy clusters \citep{jin02,fox-pen01,
suw-etal03,lee04,kas-evr05,hop-etal05,lee05,smi-wat05,hay07}, previous 
studies have been largely focused on the statistical treatment of 
cluster triaxiality. For the measurement of the gas mass fraction of galaxy 
clusters that can provide powerful constraints on the density parameter and 
the dark energy equation of state \citep{whi-etal93,lub-etal96,cen97,evr97,
coo98,lar06,fer-bla07}, however, it is necessary to deal with the individual 
clusters and their triaxial shapes.

The standard picture based on the cosmic web theory \citep{bar-etal86,
bon-etal96} explains that galaxy clusters are rare events corresponding 
to the local maxima of the initial density field and form at the dense 
nodes of the local filaments in the cosmic web through tidal interactions
 with the surrounding matter.  The tidal effect from the surrounding matter 
results in the deviation of cluster shapes from spherical symmetry as well 
as the preferential alignments of cluster galaxies (or halos) with the 
major axes of their host clusters \citep{bin82,str-pee85,hop-etal05,kas-evr05,
lee05,alg-etal06,alt-etal06,paz06}.

In the frame of this standard scenario, \citet[][hereafter, LK06]{lee-kan06} 
have recently developed an analytic algorithm to reconstruct the triaxial 
shapes of individual clusters. The key concept of the LK06 algorithm is that 
the two axial ratios of a triaxial cluster are related to the eigenvalues of 
the local tidal shear tensor. By measuring the spatial alignment of the 
cluster galaxies with the major axis of their host cluster, one can 
determine the eigenvalues of the local tidal tensor, which will in turn 
yields the two axial ratios of a triaxial cluster. 

Testing their analytic model against high-resolution N-body simulations, 
LK06 have shown that their algorithm works well within $20\%$ errors. 
Now that the LK06 algorithm is known to work in principle, it is time to 
test the algorithm against observations. Our goal here is to apply the 
LK06 algorithm to real observational data and examine its validity in 
practice. Here, we use the Virgo cluster as a target, whose triaxial 
shape has been very recently measured observationally \citep{mei-etal07}.  

The organization of this paper is as follows. In \S 2, a brief overview of 
the LK06 algorithm is provided and how to incorporate the two dimensional 
projection effect into the algorithm is explained. In \S 3, the Virgo cluster 
data are analyzed and its triaxial shapes are reconstructed using the LK06 
algorithm. In \S 4, the results are summarized, and the advantages and the 
caveats of our model are discussed.
  
\section{THEORETICAL MODEL}

\subsection{\it Overview of the LK06 Algorithm}

According to the LK06 algorithm, the cluster triaxial shape originates from 
its tidal interaction with the surrounding matter distribution.
This assumption has been verified from high-resolution N-body simulation which 
demonstrated clearly that the tidal field elongates the cluster shapes
\citep[e.g.,][and references therein]{alt-etal06}. LK06 has shown that the 
two axial ratios of a triaxial cluster are related to the three eigenvalues 
of the local tidal tensor defined as the second derivative of the 
gravitational potential: 
\begin{equation}
\label{eqn:axi}
\frac{b}{c}=\left (\frac{1-\lambda_{2}}{1-\lambda_{3}}\right)^{1/2} , \qquad
\frac{a}{c}=\left (\frac{1-\lambda_{1}}{1-\lambda_{3}}\right)^{1/2} ,
\end{equation}
where $\{a,b,c\}$ (with $a \le b \le c$)are the three principal axis lengths 
of a cluster and $\{\lambda_{1},\lambda_{2},\lambda_{3}\}$ (with 
$\lambda_{1} \ge \lambda_{2} \ge \lambda_{3}$) are the three eigenvalues of 
the local tidal tensor, ${\bf T}$.  According to this formula, one can 
estimate the cluster axial ratios if $\lambda_{1},\lambda_{2},\lambda_{3}$ 
is fixed by fitting the probability density distribution, $p(\cos\theta)$ 
analytically to the observational data. LK06 suggested that the preferential 
locations of the cluster galaxies near the cluster major axes given the local 
tidal tensor be described as 
\begin{equation}
\label{eqn:corr}
\langle \hat{x}_i\hat{x}_j\vert\hat{T}\rangle = 
\frac{1-s}{3}\delta_{ij}+s\hat{T}_{ik}\hat{T}_{kj}.
\end{equation}
where $\hat{\bf x} \equiv (\hat{x}_{i})$ is the unit position vector of a 
cluster galaxy, $\hat{\bf T}\equiv {\bf T}/\vert{\bf T}\vert$ is the 
unit tidal shear tensor, and $s \in [-1,1]$ is the correlation parameter 
that represents the correlation strength between $\hat{\bf x}$ and
 $\hat{\bf T}$. Under the assumption that the minor axis of the tidal shear 
tensor is in the direction of the cluster major axis, 
equation (\ref{eqn:corr}) basically describes the alignment between the 
position of a cluster galaxy and the major axis of its host cluster.  
If $s =-1$, there is a maximum alignment. If $s=1$, there is a maximum 
anti-alignment.  The case of $s=0$ corresponds to no alignment.

Let $\theta_{3d}$ be the angle between the host cluster major axis and the 
galaxy position vector. Under the assumption that the cluster major axis is 
in the direction of the minor principal axis of the tidal shear tensor,   
The probability density distribution of $\cos\theta_{3d}$ was derived by 
LK6 as 
\begin{equation}
\label{eqn:gen}
p(\cos\theta_{3d})=\frac{\vert M\vert}{2\pi}
(\hat{x}_{i}\cdot M^{-1}_{ij}\cdot\hat{x}_{j})^{-\frac{3}{2}},
\end{equation}
where $\phi$ is an azimuthal angle of $\hat{\bf x}$ measured in an arbitrary 
coordinate system. Here, covariance matrix $\bf{M}$ is defined as 
${\bf M}\equiv \langle\hat{x}_i\hat{x}_j\vert\hat{T}\rangle$. Note that 
equation (\ref{eqn:gen}) holds good for any arbitrary coordinate 
system in which the tidal shear tensor is not necessarily diagonal. 

In the principal axis frame of the tidal tensor, equation (\ref{eqn:pri}) 
can be expressed only in terms of the three eigenvalues of the tidal shear 
tensor as  
\begin{eqnarray}
 \label{eqn:pri}
p(\cos\theta)&=&\frac{1}{2\pi} \prod_{i=1}^{3}
(1-s+3s\hat{\lambda}^{2}_{i})^{-1/2}\nonumber \\
&&\times \int_{0}^{2\pi}\left(\frac{\sin^2\theta  \cos^2\phi}{1-s+3s
 \hat{\lambda}^{2}_{1}} +\frac{\sin^2\theta \sin^2\phi}{1-s+3s
 \hat{\lambda}^{2}_{2}}+\frac{\cos^2\theta}{1-s+3s\hat{\lambda}^{2}_{3}} 
  \right)^{-3/2} d\phi,
\end{eqnarray}
where $\{\hat{\lambda}_{i}\}^{3}_{i=1}$ are the eigenvalues of $\hat{\bf T}$, 
related to $\{\lambda_{i}\}^{3}_{i=1}$ as
 \begin{equation}
  \label{lambda}
  \lambda_{1}=\frac{\delta_{c}\hat{\lambda}_{1}}{\hat{\lambda}_{1}
  +\hat{\lambda}_{2}+\hat{\lambda}_{3}} ,\quad \lambda_{2}=
\frac{\delta_{c}
   \hat{\lambda}_{2}}{\hat{\lambda}_{1}+\hat{\lambda}_{2}+
   \hat{\lambda}_{3}},\quad \lambda_{3}=\frac{\delta_{c}\hat{\lambda}_{3}}
   {\hat{\lambda}_{1}+\hat{\lambda}_{2}+\hat{\lambda}_{3}},
  \end{equation}
where $\delta_{c} \approx 1.68 $ is the linear density threshold for a 
dark halo \citep{eke96} satisfying the following constraint 
of $\delta_{c}=\sum_{i=1}^{3}\lambda_{i}$.

The key concept of LK06 algorithm is that by fitting equation (\ref{eqn:gen}) 
to the observed probability distribution, one can find the best-fit values 
of $\lambda_{1},\lambda_{2}$ and $s$, and then determine the cluster axial 
ratios using equation (\ref{eqn:axi}).  Although LK06 algorithm allows us 
in principle to reconstruct the three dimensional intrinsic triaxial 
shapes of individual clusters, it is restricted to the cases where the 
informations on the three dimensional positions of the cluster galaxies 
in the cluster principal axis frame are given. Unfortunately, for most 
clusters, these informations are not available but only two dimensional 
projected images of clusters.

In $\S 2.2$, we attempt to modify the LK06 algorithm in order to incorporate 
the two dimensional projection effect. 
 
\subsection{\it Projection Effect}

Let us suppose that the position vectors of the cluster galaxies 
are all projected along the line of sight direction onto the plane of a sky. 
Unless the major axis of the host cluster is perfectly aligned with the 
line-of-sight direction to the cluster center, one would expect that the 
projected position vectors of the cluster galaxies should show a tendency 
to be aligned with the projected major principal axes. 

Let $\theta_{2d}$ be the angle between the projected cluster major axis and 
galaxy position vector in the plane of sky. The probability distribution can  
be calculated by integrating equation (\ref{eqn:gen}) along the line of sight 
as
\begin{equation}
\label{eqn:2dp}
p(\cos\theta_{2d})=\int_{-1}^{1}\frac{\vert M\vert}{2\pi}
(\hat{x}_{i}\cdot M^{-1}_{ij}\cdot\hat{x}_{j})^{-\frac{3}{2}}d
\hat{x}_{3},
\end{equation}
where $(\hat{x}_{3})$ is now chosen to be in the direction of the line of 
sight. In other words, we consider a certain Cartesian coordinate system 
in which the $\hat{x}_{3}$ direction is parallel to the line of the sight 
to the cluster center of mass. Note that in this coordinate system the 
tidal tensor is not necessarily diagonal. 

Through the similarity transformation 
\begin{equation}
\label{eqn:rot}
\hat{\bf T} =\bf{R}^{t}\cdot\hat{\bf \Lambda}_{\rm T} 
\cdot {\bf {R}}, 
\end{equation}
with  
\begin{equation}
\label{eqn:dia}
\mathbf{\Lambda_{T}} = 
\left( \begin{array}{ccc}
\hat{\lambda_1} & 0 & 0 \\
 0 & \hat{\lambda_2} & 0 \\
   0    &  0 & \hat{\lambda_3}     
\end{array} \right), 
\end{equation}
one can express the nondiagonal unit tidal tensor $\hat{\bf T}$ in 
terms of its eigenvalues.Here, the rotation matrix ${\bf R}$ has the form 
of \citep{bin85}
\begin{equation}
\label{eqn:bin}
\mathbf{R} = 
\left( \begin{array}{ccc}
-\sin\psi & -\cos\psi\cos\xi & \cos\psi\sin\xi \\
 \cos\psi & -\sin\psi\cos\xi & \sin\psi\sin\xi \\
   0    &    \sin\xi   &    \cos\xi     
\end{array} \right),
\end{equation}
where $(\xi,\psi) $ is the polar coordinate of the line-of-sight 
direction in the principal axis frame of the cluster. 

Through equations (\ref{eqn:2dp})-(\ref{eqn:bin}), we finally final 
an analytic expression for the probability density  distribution of 
$\cos\theta{2d}$ in terms of $\{\hat{\lambda}_{i}\}_{i=1}^{3}$ and 
the correlation parameter $s$: 
\begin{equation}
\label{real}
p(\cos\theta_{2d})=\frac{(1+s)^2(1-2s)}{2\pi}\int_{-1}^{1}Q^{-3/2}d\hat{x}_3,
\end{equation} 
with the factor $Q$ defined as
\begin{eqnarray}
\label{xi}
Q &\equiv & [(1 + s)^2 - 3s(1 + s)](A_1\hat{\lambda}_{1}+A_2\hat{\lambda}_{2}
        +A_3\hat{\lambda}_{3})(1 - \hat{x}^{2}_{3}), \nonumber \\
   && + [(1 + s)^2 - 3s(1 + s)](C_1\hat{\lambda}_{1} + C_2\hat{\lambda}_{2} 
+ C_3\hat{\lambda}_{3})\hat{x}^{2}_{3},
\nonumber \\
   && +  6s(1 + s)(B_1\hat{\lambda}_{1}+ B_2\hat{\lambda}_{2} + 
B_3\hat{\lambda}_{3})\hat{x}_{3}\sqrt{1 - \hat{x}^{2}_{3}},
\end{eqnarray}
where the coefficients $\{A_{i}\}^{3}_{i=1},\{B_{i}\}^{3}_{i=1},
\{C_{i}\}^{3}_{i=1}$ are given as
\begin{eqnarray}
\label{a}
A_1 &=& 
\cos\psi\cos\theta_{2d}(\cos\psi\cos\theta_{2d} -
2\sin\psi\cos\xi\sin\theta_{2d}) +
     \sin^2\theta_{2d}(\sin^2\psi + \sin^2\xi\cos^2\psi), \cr
A_2 &=& 
\sin\psi\cos\theta_{2d}(\sin\psi\cos\theta_{2d} + 
2\cos\psi\cos\xi\sin\theta_{2d}) +
     \sin^2\theta_{2d}(\cos^2\psi + \sin^2\xi\sin^2\psi), \cr
A_3 &=& \cos^2\theta_{2d} + \cos^2\xi\sin^2\theta_{2d}, \cr
B_1 &=& -\sin\xi\cos\psi(\cos\xi\sin\theta_{2d}\cos\psi + 
\sin\psi\cos\theta_{2d}), \cr
B_2 &=& -\sin\xi\sin\psi(\cos\xi\sin\theta_{2d}\sin\psi - 
\cos\psi\cos\theta_{2d}), \cr
B_3 &=& \cos\xi\sin\xi\sin\theta_{2d}, \cr
C_1 &=& \sin^2\psi + \cos^2\xi\cos^2\psi, \nonumber \cr
C_2 &=& \cos^2\psi + \cos^2\xi\sin^2\psi, \nonumber \cr
C_3 &=& \sin^2\xi .
\end{eqnarray}
With this new modified algorithm in the two dimensional space, we can 
reconstruct the intrinsic shape of a triaxial cluster halo from the 
observed two dimensional image. 

\section{APPLICATION TO THE VIRGO CLUSTER HALO}

\subsection{Observational Data and Analysis}

The Virgo cluster is the nearest richly populated cluster of galaxies 
whose properties has been studied fruitfully for long \citep[e.g.,]
[and references therein]{boh94,wes00}. It is known to have 
approximately $1275$ member galaxies \citep{bin-etal85} and 
located at a distance of approximately $16.1$ Mpc from us \citep{ton01} 
with the major axis inclined at an angle of approximately $10^{o}$ with 
respect to the line of sight \citep{wes00}.  

Near the center of the Virgo cluster is located the large ellipticity galaxy 
$M87$ (or Virgo A). The major axis of the Virgo cluster is found to be 
inclined approximately $10^{\circ}$ with respect to the line of sight 
direction to $M87$ \citep{wes00}. We use data from the Virgo Cluster 
Catalogue \citep{bin-etal85} which compiles the equatorial coordinates of 
total $1275$ member galaxies.

\subsection{Coordinate Transformation and Projection Effect}

Let $r$ and $(\alpha,\delta)$ represent the three dimensional distance to a 
member galaxy and its equatorial coordinates, respectively. 
Under the assumption that the position of $M87$ is the center of mass of the 
Virgo cluster, the Cartesian coordinate of a member galaxy in the center of 
mass frame can be written as
\begin{eqnarray}
\label{eqn:flat}
x_1 &=&r\cos\alpha\cos\delta - x_{\rm 1va} \nonumber \\
x_2 &=&r\sin\alpha\cos\delta - x_{\rm 2va} \nonumber \\
x_3 &=&r\sin\delta - x_{\rm 3va} 
\end{eqnarray}
where $(x_{\rm 1va},x_{\rm 2va},x_{\rm 3va})$ represents the position of 
$M87$. The equatorial coordinates of $M87$ is measured to be 
$\alpha_{\rm va}=187.71^{\circ}$ and $\delta_{\rm va}=12.39^{\circ}$.  
The distance to Virgo A, $r_{va}$,  from us is also known to be 
approximately $16.1$Mpc (\citealt{ton01}; SBF survey). 
Thus,we have a full information on $(x_{\rm 1va},x_{\rm 2va},x_{\rm 3va})$.

Now let us consider a coordinate system where the third axis is in 
in the direction of the line of sight to $M87$. Let $(u,\vartheta,\varphi)$
be the spherical polar coordinate of the member galaxy in this coordinate 
system. It can be found through coordinate transformation as  
\begin{equation}
 \label{eqn:tra}
\left(
\begin{array}{c} u \\ \vartheta \\ \varphi 
\end{array}\right)=
 \left(
 \begin{array}{ccc}
 \sin\xi_{\rm va}\cos\psi_{\rm va} & \cos\xi_{\rm va}\cos\psi_{\rm va} 
& -\sin\psi_{\rm va} \\ 
 \sin\xi_{\rm va}\sin\psi_{\rm va} & \cos\xi_{\rm va}\sin\psi_{\rm va} 
& \cos\psi_{\rm va} \\
  \cos\xi_{\rm va} & - \sin\xi_{\rm va}  & 0 
 \end{array} \right) 
  \left(
 \begin{array}{c} x_{1}\\ x_{2} \\ x_{3} 
 \end{array}\right ) 
  \end{equation}
where $(\xi_{\rm va},~\psi_{\rm va})$ is the polar coordinate of M87, 
so that $\xi_{\rm va}\equiv \pi/2-\delta_{\rm va}$. 

In the plane of sky projected along the line of sight direction, the 
spherical polar coordinates of the member galaxy can be regarded as 
the Cartesian coordinates as $r \to 0$, $\vartheta \to x_{l}$, and 
$\phi \to y_{l}$. Basically, it represents a two dimensional projected 
position of a Virgo cluster member galaxy in the plane of sky with the 
position of $M87$ as a center.

For each member galaxy from the Virgo Cluster Catalogue, we have determined 
the two dimensional projected position $(x_{l},y_{l})$ using the given 
equatorial coordinates. 

\subsection{Virgo Cluster Reconstruction}

To measure the alignments between the positions of the Virgo cluster galaxies 
and the projected major axis of the Virgo cluster and compare the distribution 
of the alignment angles with the analytic model (eq.[\ref{eqn:2dp}]), it is 
necessary to find the direction of the projected major axis in the coordinate 
system of $(x_{1l},x_{2l})$. Equivalently, it is necessary to have an 
information on the polar coordinates of the line of sight, $(\xi,\psi)$ 
with respect to the three dimensional principal axis of the Virgo cluster.

The seminal paper of \citet{wes00} provides us with the information on 
$\xi$ i.e., the angle between the three dimensional major axis of the Virgo 
cluster and the line of sight direction as approximately $10^{\circ}$. 
However, we still need the azimuthal angle $\psi$ of the major axis of the 
Virgo cluster. 

To find the azimuthal angle $\psi$, we first let $\xi=10^{\circ}$ and 
$\psi=0$ in the analytic model (eq.[\ref{eqn:2dp}]). It implies that we 
choose a certain Cartesian coordinates $(x_{1p},x_{2p})$ where the angle 
$\psi$ vanishes. Then, we transform the coordinate system 
of $(x_{1l},x_{2l})$ into this new coordinate system as 
\begin{equation}
 \label{eqn:new}
 \left(\begin{array}{cc} x_{1p} \\ x_{2p} \end{array}\right) 
 =\left(\begin{array}{cc} \cos\psi & \sin\psi \\ -\sin\psi &
 \cos\psi \end{array}\right) \left(\begin{array}{cc} x_{1l} \\  x_{2l} 
 \end{array}\right).
 \end{equation}
Here, note that $(\cos\psi, \sin\psi)$ corresponds to the projected 
major axis of the Virgo cluster in the $(x_{1l},x_{2l})$ coordinate system.
It is expected that in this new $(x_{1p},x_{2p})$-coordinate system the 
the observationally measured distribution should fit the analytic model 
(eq.[\ref{eqn:2dp}] best. 

For a given $\psi$, we measure the alignment angle between the projected 
major axis and position vector of each Virgo cluster galaxy as 
\begin{eqnarray}
\label{cos}
\cos\theta_{2d} = \hat{x}_{1p}\cos\psi + \hat{x}_{2p}\sin\psi, 
\end{eqnarray}
where $\hat{\bf x}_{p}\equiv {\bf x}_{p}/\vert{\bf x}_{p}\vert$. Then, by 
counting the number of galaxies galaxy's number density as a function of 
$\cos\theta_{d}$, we can derive the probability distribution, 
$p(\cos\theta_{d})$. We fit this observational distribution with the analytic 
model, adjusting the values of $\lambda_{1},\lambda_{2}$ and $s$ through 
$\chi^{2}$-minimization. We repeart the whole process varying the value 
of $\psi$, and seek for the value of $\psi$ which yields the smallest 
$\chi^{2}$ value. As a final step, we determine the corresponding best-fit 
values of $\lambda_{1}$ and $\lambda_{2}$. 

\begin{figure}
  \begin{center}
   \plotone{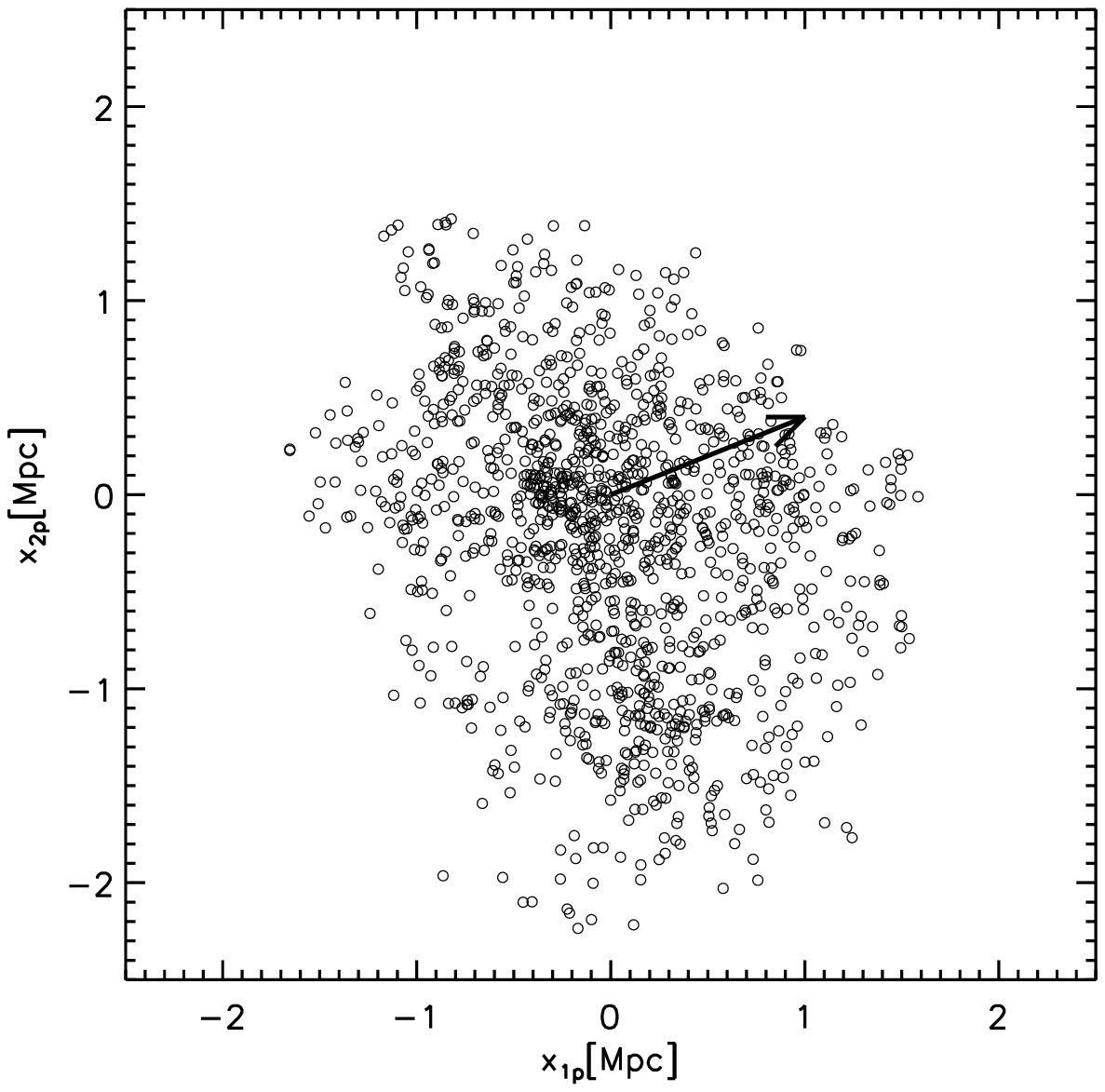}
\caption{Positions of the Virgo member galaxies in the two dimensional 
projected space. The arrow represents the direction of the projected major 
axes of the Virgo cluster.}
\label{fig:plane}
 \end{center}
\end{figure}

Finally, we find the axial ratios of the Virgo cluster to be $a/c=0.54$ 
and $b/c=0.73$ by using equation (\ref{eqn:axi}) with the constraint of 
$\delta_{c}=\sum_{i=1}^{3}\lambda_{i}$. The best-fit value of $s$ is also 
determined to be $-0.25$, indicating that the positions of the cluster 
galaxies are indeed aligned with the major axes of the Virgo cluster. 
Figure [\ref{fig:plane}] plots the two dimensional projected positions of 
the Virgo cluster galaxies in the $(x_{1p},x_{2p})$-coordinate system. 
The arrow represents the direction of the projected major axis determined 
from the chosen value of $\psi$. 

\begin{figure}
  \begin{center}
   \plotone{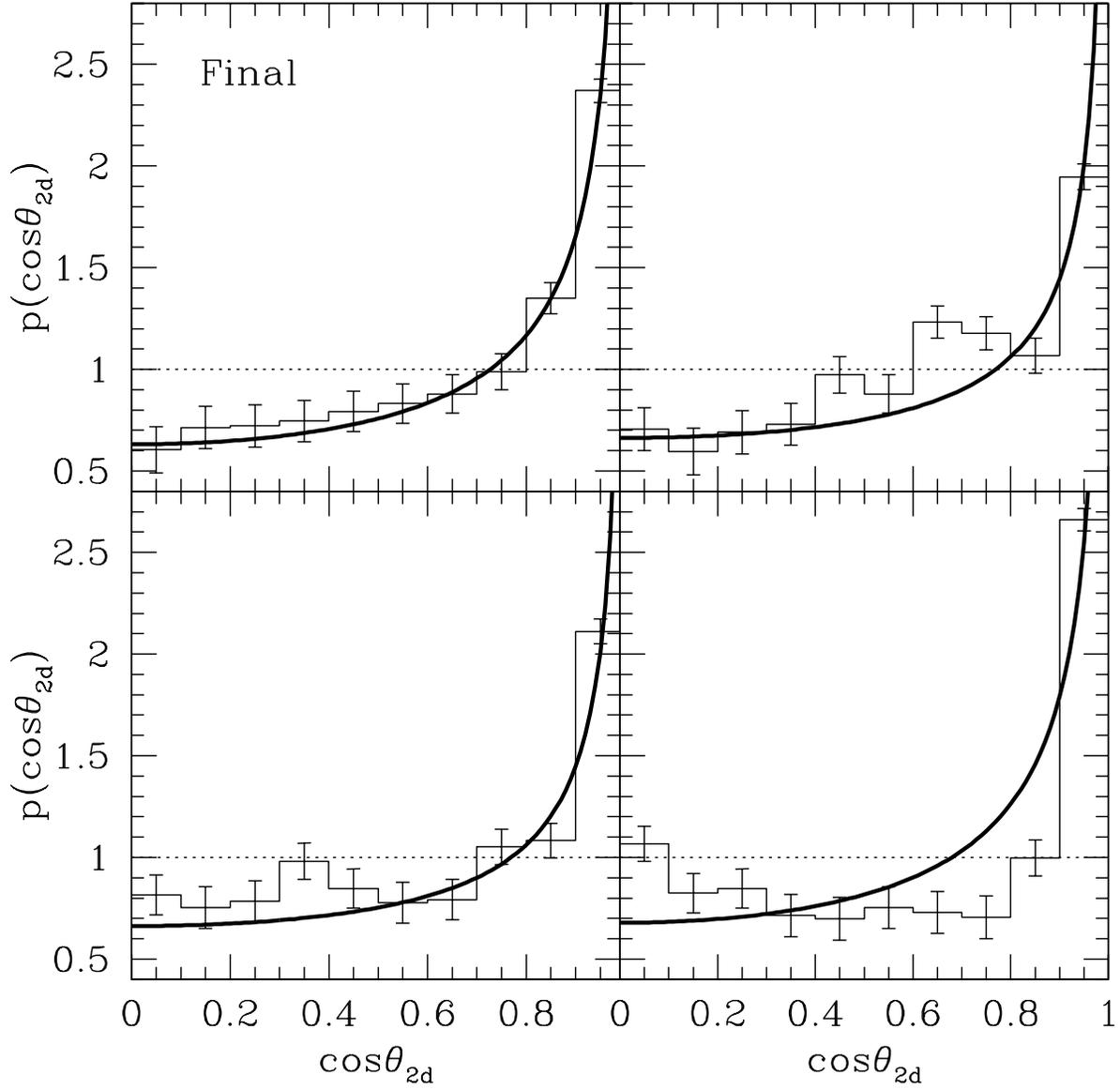}
\caption{Probability density distributions of the cosines of the angles 
between the position vectors of the Virgo cluster galaxies and four different 
choices of the Virgo cluster major axis in the two dimensional projected 
plane of sky. In each panel, the histogram with Poisson errors represents the 
observational data points from the Virgo Catalog, the solid line is the 
analytic fitting function based on the LK06 reconstruction algorithm, and 
the dotted line corresponds to the case of no correlation.
The top left panel corresponds to the best-fit result according to which 
the major axes of the Virgo cluster in the two dimensional projected plane 
is determined, while the other three panels show how the agreements between 
the observational and the analytic results change if different directions 
other than the major axes are used.}
\label{fig:fit}
 \end{center}
\end{figure}

Figure \ref{fig:fit} plots the probability distribution of the alignment 
angles for four different cases of $\psi$. In each panel, the histogram 
with Poisson errors is the observational distribution while the solid line 
represents the analytic fitting model. The dotted line stands for the 
case of no alignment at all. The top left panel corresponds to the 
finally chosen value of $\psi$ for which the analytic fitting 
model and the observational result agree with each other best. The other 
three panels show the three exemplary cases of $\psi$ for which the 
analytic fitting model and the observational result do not agree with 
each other well with relatively high $\chi^{2}$ value.

To investigate whether the reconstructed axial ratios of the Virgo cluster 
changes with the radius from the center, we introduce a cut-off radius, 
$R_{\rm cut}$, and remeasure the axial ratios using only those galaxies 
located within $R_{\rm cut}$ from the center in the $(x_{1p},x_{2p})$-
corrodinate system. We repeat the same process but using four different 
values of $R_{\rm cut}$. Table \ref{tab:fit} lists the values of the 
resulting best-fit axial ratios $a/c$ and $b/c$, and the best-fit correlation 
parameter for the four different cut-off radii of $R_{cut}$. As can be seen, 
at the inner radius smaller than the maximum one, the shape tends to be more 
prolate-like, consistent with recent numerical report \citep[e.g.,][]{hay07}. 
Note also that the value of $s$ is consistently $-0.25$, which implies the 
strength of the tidal interaction with surrounding matter is consistent.

\begin{figure}
  \begin{center}
   \plotone{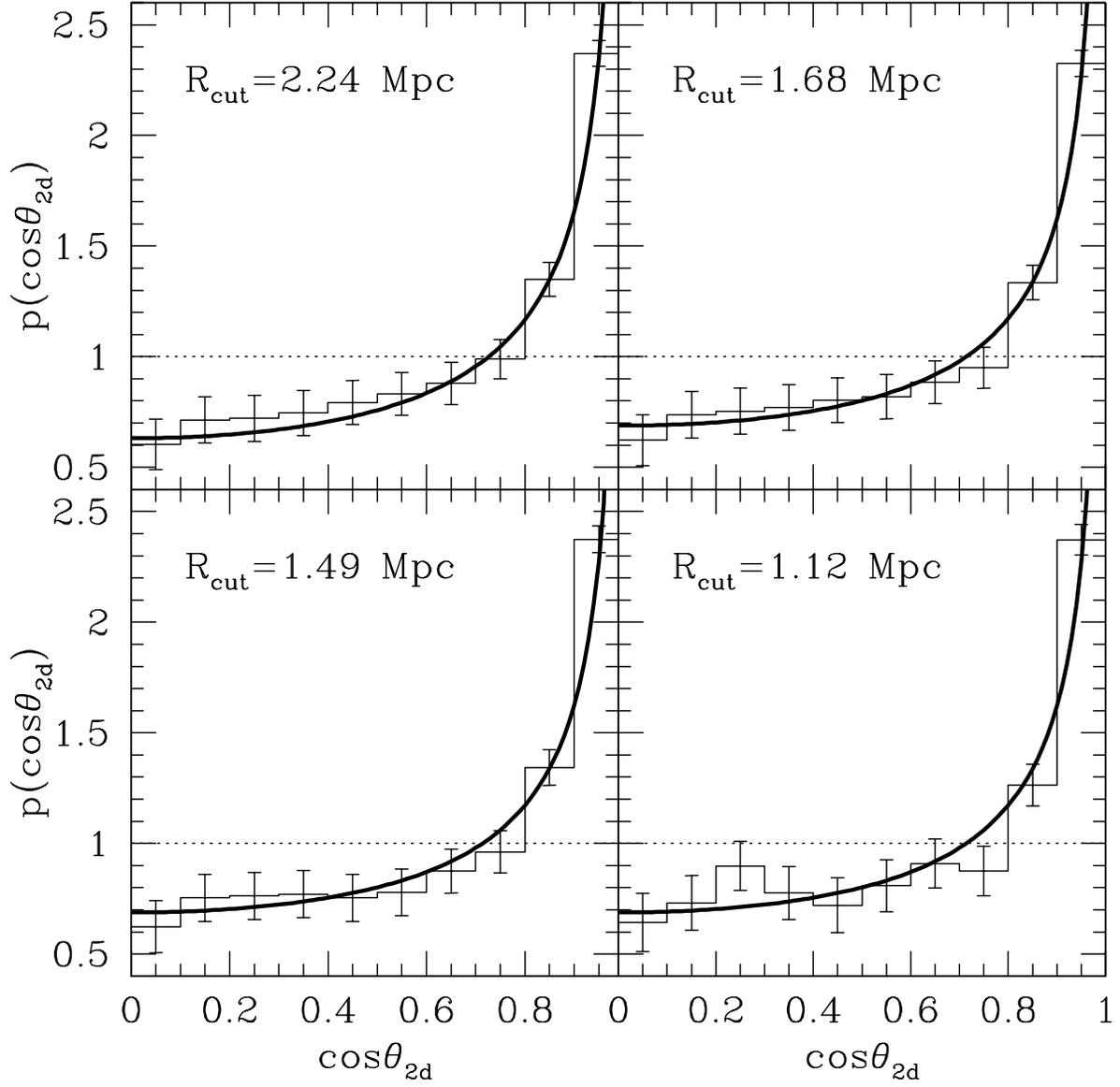}
\caption{Comparison between the observational and the analytic results for 
the alignments at different two dimensional cut-off radii ($R_{\rm cut}$).}
\label{fig:radii}
 \end{center}
\end{figure}

Figure \ref{fig:radii} plots the probability distributions for the four 
different cases of $R_{\rm cut}$. The top left panel shows the case of maximum 
cut-off radius. As can be seen, the agreements between the analytic fitting 
model and the observational result are quite good even at inner radii.

\section{SUMMARY AND DISCUSSION}

We have modified the cluster reconstruction algorithm which was originally 
developed by \citet{lee-kan06} to apply it to the two dimensional projected 
images of galaxy clusters in practice. Assuming that unless the cluster major 
axes are in the line-of-sight direction, we have found that the alignments 
between the galaxy positions and the projected major axes can be used to 
reconstruct the two axial ratios of the triaxial clusters. We have applied 
the modified algorithm to the observational data of the Virgo cluster and 
shown that the reconstructed axial ratios of ($1$: $0.54$ : $0.73$) are in 
good agreement with the recent report from the Virgo Cluster survey 
\citep{mei-etal07}, which proves the validity and usefulness of our 
method in practice.

Now that the Virgo cluster is found to be triaxial, let us discuss on 
the triaxiality effect on the mass estimation. For simplification, let us 
assume that the Virgo cluster has a uniform density. Then, the mass of 
the triaxial Virgo cluster with the axial ratios given as $(1:0.54:0.73)$ 
is estimated to be $2.6$ times larger than the spherical case since the 
spherical radius is close to the minor axis length of the Virgo cluster 
since the major axis of the Virgo cluster is very well aligned with the 
line of sight direction. Therefore, it can cause maximum $\sim 50\%$ errors 
to neglect the Virgo cluster triaxiality.  

The most prominent merit of our method over the previous one is that 
it reconstructs directly the {\it intrinsic three dimensional structures} 
of the underlying triaxial dark matter halo using the fact that the cluster 
triaxiality originated from the tidal interaction.  
Conventionally, the triaxial shape of a cluster is found through calculating 
its inertia momentum tensor. This conventional method, however, is unlikely 
to yield the {\it intrinsic} shape of the underlying dark matter halo unless 
the target cluster is a well relaxed system. In addition, our method does not 
resort to any simplified assumption like the axis-symmetry or the alignment 
with the line-of-sight and etc. 

Yet, it is worth mentioning here a couple of limitations of our method. 
First, it assumes that the major axes of the clusters are not aligned with 
the line-of-sight direction so that the alignments between the galaxy 
positions and the projected major axes can be measured. 

Second, for the Virgo cluster, the crucial information on the angle 
$\xi$ between the three dimensional major axis and the line of sight 
has been already given \cite{wes00}. Which simplifies the whole of 
our method since it was only $\psi$ that has to be determined. 
For most clusters, however, this information is not given, so that 
both the values of $\xi$ and $\psi$ have to be determined through 
fitting before finding the axial ratios. 

Third, its success is subject to the validity of the LK06 algorithm. 
According to the numerical test, the LK06 algorithm suffers approximately 
$20\%$ errors for the case that the number of the member galaxies is not 
high enough. It implies that the LK06 algorithm is definitely restricted 
to the rich clusters with large numbers of galaxies. 
Therefore, it may be necessary to refine and improve the LK06 algorithm 
itself for the application to poor cluster samples. Our future work is in 
this direction.

\acknowledgments

This work is supported by the research grant No. R01-2005-000-10610-0 from 
the Basic Research Program of the Korea Science and Engineering Foundation.

\clearpage

\clearpage
\begin{deluxetable}{ccccc}
\tablewidth{0pt}
\setlength{\tabcolsep}{5mm}
\tablehead{$R_{\rm cut}$  & $N_{g}$ & $a/c$   & $ b/c $ & $s$ \\ 
(Mpc)    &        &         &         & }
\tablecaption{The cut-off radius ($R_{\rm cut}$) in the two dimensional 
projected space, the number of the member galaxies enclosed within 
$R_{\rm cut}$, the reconstructed two axial ratios, and the best-fit value 
of the correlation parameter.}
\startdata   
$2.24$    & $1275$   & $0.53$ & $0.73$ & $-0.25$   \\ 
$1.68$    & $1221$   & $0.64$ & $0.69$ & $-0.25$   \\ 
$1.49$    & $1154$   & $0.64$ & $0.69$ & $-0.25$   \\ 
$1.12$    & $902$    & $0.64$ & $0.69$ & $-0.25$   \\ 
\enddata
\label{tab:fit}
\end{deluxetable}

\clearpage
\begin{thebibliography}{100}
\bibitem[Algood et al.(2006)]{alg-etal06}
Algood, B. et al. 2006, \mnras, 367, 1781  
\bibitem[Altay et al.(2006)]{alt-etal06}
Altay, G., Colberg, J. M., \& Croft, R. A. C. 2006, \mnras, 370, 1422
\bibitem[Anninos \& Norman(1996)]{ann96}
Anninos, P., \& Norman, M. L. 1996, \apj, 459, 12
\bibitem[Bardeen et al.(1986)]{bar-etal86}
Bardeen, B., Bond, R. J., Kaiser, N., \& Szalay, H. 1986, \apj, 23, 567
\bibitem[Basilakos et al.(2000)]{bas00}
Basilakos, S., Plionis, M., \& Maddox, S.J. 2000, \mnras, 316, 779
\bibitem[Binggeli(1982)]{bin82}
Binggeli, B. 1982, \aap, 107, 338
\bibitem[Binggeli et al.(1985)]{bin-etal85}
Binggeli, B., Sandage, A., \& Tammann, G., A. 1985, \aj, 90, 1681
\bibitem[Binney(1985)]{bin85}
Binney, J. 1985, \mnras, 212, 767
\bibitem[Bohringer et al.(1994)]{boh94}
Bohringer, H. et al. 1994, Nature, 368, 828
\bibitem[Bond et al.(1996)]{bon-etal96}
Bond, J. R., Kofman, L., \& Pogosyan, D. 1996, Nature, 380, 603
\bibitem[Cen(1997)]{cen97}
Cen, R. 1997, \apj, 485, 39
\bibitem[Cooray(1998)]{coo98}
Cooray, A. R. 1998, \aap, 339, 623
\bibitem[Eke et al.(1996)]{eke96}
Eke, V. R., Cole, S., \& Frenk, C. S. 1996, \mnras, 282, 263
\bibitem[Evrard(1997)]{evr97}
Evrard, A. E. 1997, \mnras, 292, 289
\bibitem[Fabricant \& Gorenstein(1983)]{fab83}
Fabricant, D., Gorenstein, P. 1983, \apj, 267, 535
\bibitem[Ferramacho \& Blanchard(2007)]{fer-bla07}
Ferramacho, L. D., \& Blanchard, A. 2007, \aap, 463, 423
\bibitem[Fouque et al.(2001)]{fou01}
Fouque, P. S., Solanes, J. M., Sanchis, T., \& Balowski, C. 2001, 
\aap, 375, 770
\bibitem[Fox \& Pen(2001)]{fox-pen01}
Fox, D. C., \& Pen U. L. 2002, \apj, 574, 38
\bibitem[Frenk et al.(1988)]{fre-etal88}
Frenk, C. S., White, S. D. M., Davis, M., Efstathiou, G. 1988, \apj, 327, 507
\bibitem[Hayashi et al.(2007)]{hay07}
Hayashi, E., Navarro, J. F., \& Springel, V. 2007, accepted \mnras
\bibitem[Hopkins et al.(2005)]{hop-etal05}
Hopkins, P. F., Bahcall, N., \& Bode, P. 2005, \apj, 618, 1
\bibitem[Jing \& Suto(2002)]{jin02}
Jing, Y.P., \& Suto, Y. 2002, \apj, 574, 538
\bibitem[Kasun \& Evrard(2005)]{kas-evr05}
Kasun, S. F., Evrard, A. E. 2005, \apj, 629, 781
\bibitem[Klypin et al.(1999)]{kly99}
Klypin, A., Cottlober, S., \& Kravtsov, A. V. 1999, \apj, 516, 530 
\bibitem[Laroque et al.(2006)]{lar06}
Laroque, S. J. 2006, \apj, 652, 917
\bibitem[Lee \& Pen(2001)]{lee01}
Lee, J., Pen, U. L. 2001, \apj , 555, 106
\bibitem[Lee \& Suto(2003)]{lee03}
Lee, J., Suto, Y. 2003, \apj, 585, 151
\bibitem[Lee \& Suto(2004)]{lee04}
Lee, J., Suto, Y. 2004, \apj , 601, 599
\bibitem[Lee et al.(2005)]{lee05}
Lee, J., Kang, Xi, \& Jing, Y. P. 2005, \apj, 639, L5
\bibitem[Lee \& Kang (2006)]{lee-kan06}
Lee, J., Kang, Xi 2006, \apj, 637, 561
\bibitem[Lokas \& Mamon(2003)]{lok03}
Lokas, E. L., Mamon, G. A. 2003, \mnras, 343, 401
\bibitem[Lubin et al.(1996)]{lub-etal96}
Lubin, L. M., Cen, R., Bahcall, N. A., \& Ostriker, J. P. 1996, ApJ, 460, 10
\bibitem[Mamon et al.(2004)]{mam04}
Mamon, G. A., Sanchis, T. S., \& Solanes, J. M. 2004, 
\aap, 414, 445
\bibitem[Mclaughlin et al.(1999)]{mcl99}
Mclaughlin, D. E. 1999, \apj, 512, L9
\bibitem[Mei et al.(2007)]{mei-etal07}
Mei, S. et al. 2007, \apj , 655, 144 
\bibitem[Mohr et al.(1999)]{moh99}
Mohr, J. J., Mathiesen, B., \& Evrard,A. E. 1999, \apj, 517, 627
\bibitem[Nulsen \& Bohringer(1995)]{nul95}
Nulsen, P. E. J., Bohringer, H. 1995, \mnras, 274, 1093
\bibitem[Paz et al.(2006)]{paz06}
Paz, D. J., Lambas, D. G., Padilla, N., \& Merchan, M. 2006, \mnras, 366, 1503
\bibitem[Plionis et al.(1991)]{plio91}
Plionis, M., Barrow, J. D., \& Frenk, C. S. 1991, \mnras, 249, 662
\bibitem[Sanderson et al.(2003)]{san03}
Sanderson, A. J. R. 2003, \mnras, 340, 989
\bibitem[Schindler et al.(1999)]{sch99}
Schindler, S., Binggeli, B., \& Bohringer, H. 1999, \aap , 343, 420
\bibitem[Smith \& Watts(2005)]{smi-wat05}
Smith, R. E., \& Watts, P. I. R. 2005, \mnras, 360, 203
\bibitem[Solanes et al.(2002)]{sol02}
Solanes, J. M. 2002, \apj, 124, 2440
\bibitem[Struble \& Peebles(1985)]{str-pee85}
Struble, M. F., Peebles, P. J. E. 1985, \aj, 90, 582
\bibitem[Suto et al.(1998)]{sut98}
Suto, Y., Sasaki, S., \& Makino, N. 1998, \apj, 509, 544
\bibitem[Suwa et al.(2003)]{suw-etal03}
Suwa, T., Habe, A., Yoshikawa, K., \& Okamoto, T. 2003, \apj, 588, 7
\bibitem[Tonry et al.(2001)]{ton01}
Tonry, J. L. et al. 2001, \apj, 546, 681
\bibitem[Warren et al.(1992)]{war-etal92}
Warren, M. S., Quinn, P. J., Salmon, J. K., Zurek, W. H., \apj, 399, 405
\bibitem[West(1989)]{wes89}
West, M. J. 1989, \apj, 347, 610
\bibitem[West \& Blakeslee(2000)]{wes00}
West, M. J., Blakeslee, J. P. 2000, \apj, 543, L27
\bibitem[White et al.(1993)]{whi-etal93}
White, S. D. M., Navarro, J., \& Evrard, A. E., \& Frenk, C. S. 1993, 
Nature, 366, 429
\bibitem[Zentner et al.(2005)]{zen05}
Zentner, A. R. et al. 2005, \apj, 629, 219
\end{thebibliography}
\end{document}